# COMPREHENSIVE MEASUREMENT FRAMEWORK FOR ENTERPRISE ARCHITECTURES

Mahesh R. Dube[1] and Shantanu K. Dixit[2]

[1]Department of Computer Engineering, VIT, Pune, India
mrdube@rediffmail.com
[2]Department of Electronics and Telecommunications, WIT, Solapur, India
dixitsk1@yahoo.com

## ABSTRACT

*Enterprise Architecture defines the overall form and function of systems across an enterprise involving the stakeholders and providing a framework, standards and guidelines for project-specific architectures. Project-specific Architecture defines the form and function of the systems in a project or program, within the context of the enterprise as a whole with broad scope and business alignments. Application-specific Architecture defines the form and function of the applications that will be developed to realize functionality of the system with narrow scope and technical alignments. Because of the magnitude and complexity of any enterprise integration project, a major engineering and operations planning effort must be accomplished prior to any actual integration work. As the needs and the requirements vary depending on their volume, the entire enterprise problem can be broken into chunks of manageable pieces. These pieces can be implemented and tested individually with high integration effort. Therefore it becomes essential to analyze the economic and technical feasibility of realizable enterprise solution. It is difficult to migrate from one technological and business aspect to other as the enterprise evolves.*

*The existing process models in system engineering emphasize on life-cycle management and low-level activity coordination with milestone verification. Many organizations are developing enterprise architecture to provide a clear vision of how systems will support and enable their business. The paper proposes an approach for selection of suitable enterprise architecture depending on the measurement framework. The framework consists of unique combination of higher order goals, non-functional requirement support and inputs-outcomes pair evaluation. The earlier efforts in this regard were concerned about only custom scales indicating the availability of a parameter in a range.*

## 1. INTRODUCTION

American National Standards Institute/Institute of Electrical and Electronics Engineers (ANSI/ IEEE) standard 1471-2000 describes architecture as the fundamental organization of a system, embodied in its components, their relationships to each other and the environment, and the principles governing its design and evolution. Enterprise architecture is the set of representations required to describe a system or enterprise regarding its construction, maintenance and evolution. Enterprise architecture aims at creating an environment suitable for mapping the organizational assets to business processes which can identify relevance and realm of business strategy adopted. An Enterprise architecture framework typically consists of business architecture, information architecture, application system architecture, and infra-structure technology architecture.

Architecture frameworks are evaluated on the basis of scope, architecture process, verification support, standards compliance and overall complexity of the architecture. Enterprise architectures should support the business processes and indicate the benefits earned by its application. Feature extraction and enhancement are the major issues while dealing with architecture flexibility and scalability. Productivity, cost-effectiveness and optimization in terms of services are the other broad parameters affecting deployment of enterprise architectures.

It is necessary to observe the pattern of migration from platform-independent and platform-specific elements in Enterprise architecture evolution. In this paper, we are limiting the scope of views and its correspondence to The Open Group Architecture Framework (TOGAF), Generalized Enterprise Reference Architecture and Methodology (GERAM), IEEE Std 1471-2000 IEEE Recommended Practice for Architectural Description, Model-Driven Architecture (MDA) and ISO RM-ODP.

## 2. RELATED WORK

John Zachman developed a framework in 1987 which was based on plan-driven approach and best practices adoption that can be deployed within the development organizations to address enterprise engineering problems. It was based on maintaining information profile of function aspects as well as the management required to accomplish the development activities. The prime issue addressed by Zachman's framework was architecture integration and implementation with a well-designed organization structure. Cap Gemini Ernst & Young developed an approach for analysis and development of enterprise and project-level architectures known as the Integrated Architecture Framework (IAF). IAF was the first implementation of enterprise engineering solutions which was widely accepted by technical community. Similar to Zachman's framework, IAF also aims at partitioning the problem in to manageable pieces based on the area of concern. IAF starts at Business Management aspect primarily dealing with business process and taskforce management. It maps the technology problem to information as knowledge-base, Information System used for traceability, and Technology Infrastructure, with special emphasis on Security aspects and Governance. Enterprise Architecture Planning (EAP) defines a process that emphasizes techniques for organizing and directing enterprise architecture projects, obtaining stakeholder commitment, presenting the plan to stakeholders, and leading the organization through the transition from planning to implementation [1].

Federal Enterprise Architecture Framework (FEAF) was developed in 1998 with the vision of integrating federal architectural segments. The FEAF was based on knowledge and asset management across the organization with a uniform terminology used for architectural integration. The business-driven aspect of FEAF was designed in view of accommodating the current as well as future business needs. The business information was later used in planning and implementation business operations in order to realize the Enterprise Architecture. FEAF emphasized on Architecture Evolution management with the help of transitional and transformational processes [8].

The Open Group Architecture Framework (TOGAF) was based on Application Lifecycle Management which largely covered the areas of governance as applicable to related areas of problems spanning from data to security. TOGAF is considered to be as a major contribution for enterprise architecture development because of the flexibility offered as well as verification-validation support provided. The Open Group is a vendor-neutral and technology-neutral consortium seeking to enable access to integrated information, within and among enterprises, based on open standards and global interoperability [2].

The IFAC/IFIP Task Force on Architectures for Enterprise Integration developed an overall definition of a generalized architecture which focused on modeling and tools that can be used for enterprise development. Generalised Enterprise Reference Architecture and Methodology (GERAM) addressed the issue of single enterprise development as well as networked enterprise development through various views which can be used at various levels of details depending on area of specialization of the enterprise. GERAM was based on Entity oriented strategy used for enterprise development [12].

The Object Management Group (OMG) introduced the Model-Driven Architecture (MDA) initiative as an approach for system development based on specification and interoperability expressed in terms of formal models. In MDA, Platform-Independent Models (PIMs) are used to represent the target system analysis and design expressed in a general-purpose modeling language, such as Unified Modeling Language (UML). The platform-independent model can be mapped to a Platform-Specific Model (PSM) by mapping the PIM to some implementation language using set of transformational rules. The MDA considers Metamodeling as a key concept for artifact generation at all stages evolution. MDA support evolution with the help of consistent mapping of resources at source to target with the help of metamodel at the two ends as well as transformation rules along with model merging [13].

The OMG MDA comprises CWM, UML, MOF and XMI as standards for model-driven development. The Common Warehouse Metamodel (CWM) defines a metamodel representing both the business and technical metadata which can be found in the data warehousing and business analysis domains. It is used as the basis for interchanging instances of metadata between heterogeneous, multi-vendor software systems. UML, which is a general purpose modelling language provides support for modelling structural and behavioural properties of the system and is part of CWM. UML is an integrated effort of three object-oriented methods (Booch, OMT, and OOSE). UML has extensive support for modelling generic systems. UML 2.0 is widely used in reactive systems behaviour analysis. The Meta Object Facility (MOF) is an OMG standard defining a common, abstract language for the specification of metamodels. It defines the four-level structure used to represent the details of how the notation repository can be made available to the modeller on model space. MOF semantics defines metadata repository that support model construction. It has the support for applying the transformations based on metamodel level selected. XML Metadata Interchange (XMI) defines XML tags that can be used to represent objects and their associations [3] [4].

## 3. ZACHMAN FRAMEWORK

The Zachman Framework for Enterprise Architecture is a widely used and accepted approach for developing or documenting an enterprise-wide architecture. It is based on Information System Architecture (ISA) and typically used in a development environment which supports organization structures and practices [5]. It is considered to be the basis for the emergence of other eminent enterprise architectures. ZF's key goals are for enterprise architecture analysis and modelling and it is concerned with *perspective*s of constructing an information system. The Zachman Framework organized as a table as indicated in Table 1.

The rows are as follows:
- Scope: It is an executive summary for a planner.
- Business model: It indicates the business process engineering efforts and activities planned in order to achieve business goals.

- System model: It indicates data elements and software functions that represent the business model.
- Technology model: It describes the constraints of tools, technology, and materials.
- Components: It indicates smallest pieces of system that can found to be functional, tested and verified according to specification.
- Working system: It depicts the operational system.

The columns are as follows:

- Who: Represents the individuals who have enactment of fulfilment of some service.
- When: Represents achievement of explicitly stated goals or objectives by the individuals on a time line indicating activity arrival and exit.
- Why: Describes the motivations of the enterprise.
- What: Describes the activities involved in corresponding area of the enterprise.
- How: Shows the functions within each perspective.
- Where: Shows locations and interconnections within the enterprise

Table 1: Zachman Framework

|  | **Data (what)** | **Function (how)** | **Network (where)** | **People (who)** | **Time (when)** | **Motivation (why)** |
|---|---|---|---|---|---|---|
| **Scope (Planner)** | List of things important to business | List of processes the business performs | List of locations where business operates | List of organisations / agents that are important | List of significant events | List of business goals / strategies |
| **Enterprise Model (Owner)** | Semantic Model | Business Process Model | Business Logistic System | Work Flow Model | Master Schedule | Business Plan |
| **System Model (Designer)** | Logical Data Model | Application Architecture | Distributed System Architecture | Human Interface Architecture | Processing Structure | Business Rules |
| **Technology Model (Builder)** | Physical Data Model | Systems Design | Technology Architecture | Presentation Architecture | Control Structure | Rule Design |
| **Components (Subcontractor)** | Data Definition | Program | Network Architecture | Security Architecture | Timing Definition | Rule Specification |

## 4. ISO RM-ODP

The Reference Model of Open Distributed Processing (ISO-RM-ODP) provides a framework for the development of systems that supports processing under heterogeneous platforms [6]. To model distributed systems, Object-modeling approach is used in RM-ODP. RM-ODP is a joint effort by the International Organization for Standardization (ISO), the International Electrotechnical Commission (IEC) and the Telecommunication Standardization Sector (ITU-T). The problem-solution pairing can be done by the "viewpoints" which provide a way of describing the system; and the "transparencies" that identify specific problems unique to distributed systems as indicated in Figure 1. [7].

RM-ODP consists of four basic International Standards:

- Overview: It describes the overview of the ODP, Scope and terminology involved in overall architecture development.
- Foundations: It describes the significant issues and factors which should be considered for distributed processing functions and systems. .
- Architecture: It represents the characteristics possessed by distributed processing system under constraints mentioned in specification. It also recommends the use of viewpoints that can be used for logical grouping of related areas of the enterprise.
- Architectural Semantics: It focuses on the modelling with the help of formal specification techniques with adequate details of each concerned area.

The viewpoints in RM-ODP are:

- Enterprise viewpoint: It deals with the strategy that can be used to accomplish the business goals and needs as identified in the preliminary phase of problem investigation.
- Information viewpoint: It focuses on information structure, information flow, logical and physical organization of information with information change tracking.
- Computational viewpoint: It focuses on structural elements of the system and their dynamics guided by protocols represented by interfaces and functionality by objects.
- Engineering viewpoint: It indicates overall organization of the objects identified and their participation in various interaction patterns to satisfy a service.
- Technology viewpoint: It indicates hardware and software components that formulate the system.

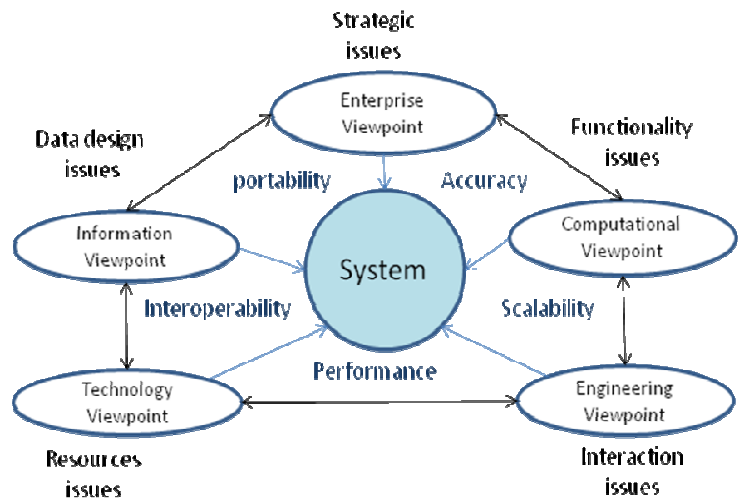

Figure 1: Viewpoints in RM-ODP

## 5. FEDERAL ENTERPRISE ARCHITECTURE FRAMEWORK (FEAF)

The goal of FEA is to improve interoperability within U.S. government agencies by creating single enterprise architecture for the entire federal government [8].

International Journal of Computer Science & Information Technology (IJCSIT) Vol 3, No 4, August 2011

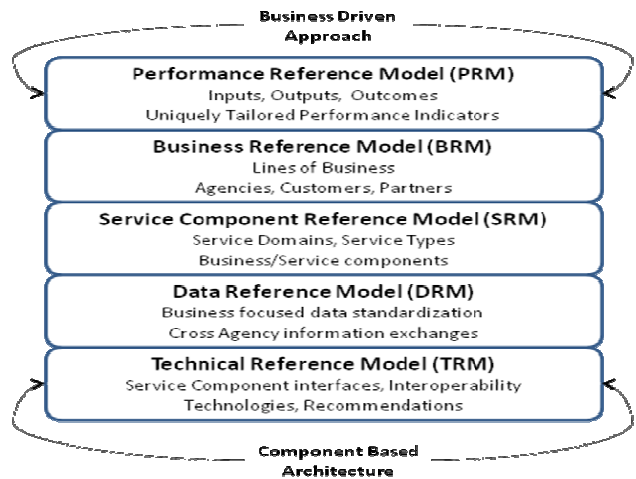

Figure 2: FEAF Reference Models

The intent of the FEAF is to enable the federal government to define and align its business functions and supporting IT systems through a common set of reference models. Figure 2 indicates FEAF Reference Models which are defined as follows:

- Performance Reference Model (PRM): The PRM is a standardized framework to measure the economics of investments and adherence to program portfolios in future based on performance.
- Business Reference Model (BRM): The BRM is a function-driven framework for describing business operations of the federal government independent of the agencies that perform them.
- Service Component Reference Model (SRM): The SRM is a framework which supports enactment of service-component relationship on the basis of performance objectives.
- Data Reference Model (DRM): The DRM is a generic model which describes the information necessary to trace operation level details.
- Technical Reference Model (TRM): The TRM is a technical framework which verifies and validates the components capabilities in relation to the specification stated and acceptable performance with reference to standards agreed upon.

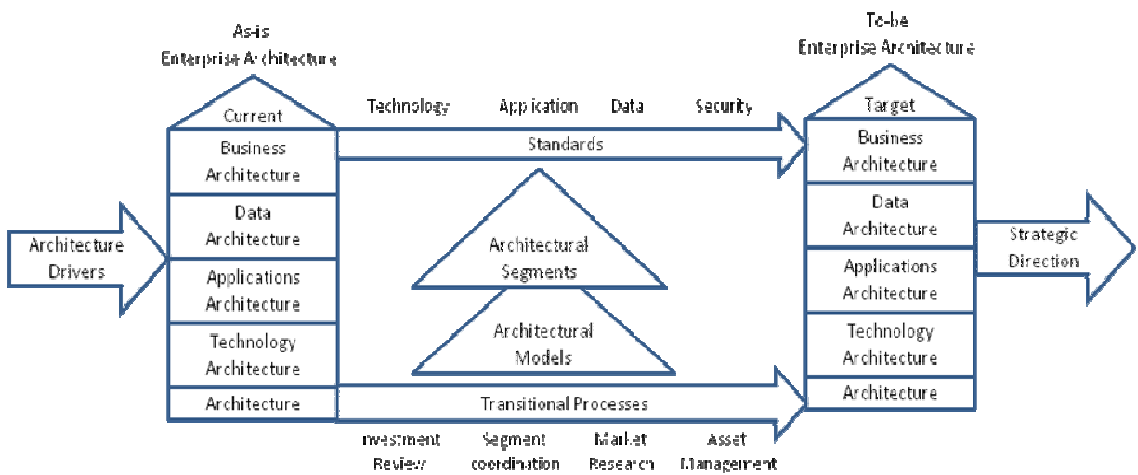

Figure 3: Simplified FEAF structure.

The major components of the FEAF are (Figure 3):
- Architecture Drivers: It indicates the factors and conditions due to which the business scenario or target design can change over a time period.
- Strategic Direction: It consists of the vision and strategic information regarding objectives to be achieved by the target architecture. The strategic direction becomes necessary to have a pilot estimate of operational effort required to realize the enterprise solution.
- Current Architecture: It defines the "as is" scenario of the enterprise architecture and consists of existing solutions to the problem identified. It describes the capabilities needed to be addressed in accordance with the limitations of the existing solution.
- Target Architecture: It defines the "to-be-built" scenario of the enterprise architecture and consists of improved architecture and performance. It indicates the changed business needs which are required to be fulfilled in accordance with the technology migration. The target architecture can be assessed by using performance metrics indicating adherence to specification.
- Transitional Processes: It supports the migration from the current to the target architecture.
- Architectural Segments: It consists of focused architecture efforts on major cross-cutting business areas.
- Architectural Models: It indicates both strategic and technical models that guide the enterprise solution which is feasible with formal representations.
- Standards: It refers to all standards, guidelines, and best practices.

## 6. IEEE1471-2000 STANDARD

The IEEE Recommended Practice for Architectural Description of Software-Intensive Systems (IEEE Std 1471-2000 aka ANSI/IEEE Std 1471-2000) introduces a conceptual model that integrates mission, environment, system architecture, architecture description, rationale, stakeholders, concerns, viewpoints, library viewpoint, views, and architectural models facilitating the expression, communication, evaluation, and comparison of architectures in a consistent manner [9].

Stakeholders are the one who are materially benefited from the solution development. The stakeholders have specific concerns and roles which should be carefully accounted while initiating and terminating the development activities. The customers or users may not have a complete view of acceptability of the solution. Therefore it is crucial to identify the stakeholder needs before the development can commence. A view indicates group of concerns as identified through partitioning of the system. A viewpoint defines a specific case of view related to a key aspect. A viewpoint indicates possible alternatives that can be considered while analyzing and designing the system rationally using appropriate modelling techniques [10]. The conceptual framework of IEEE 1471 is shown in Figure 4 and described as follows:
- A system has architecture.
- Architecture is described by one or more architecture descriptions.
- An architecture description is composed of one or more of stakeholders, concerns, viewpoints, views, and models.
- A stakeholder has one or more concerns.
- A concern has one or more stakeholders.

International Journal of Computer Science & Information Technology (IJCSIT) Vol 3, No 4, August 2011

- A viewpoint indicates possible alternatives for relevant stakeholders.
- A view conforms to one viewpoint.
- A viewpoint defines the reason for existence of the model.
- A view can have collective representations guiding more than one view.
- A viewpoint library is composed of viewpoints.

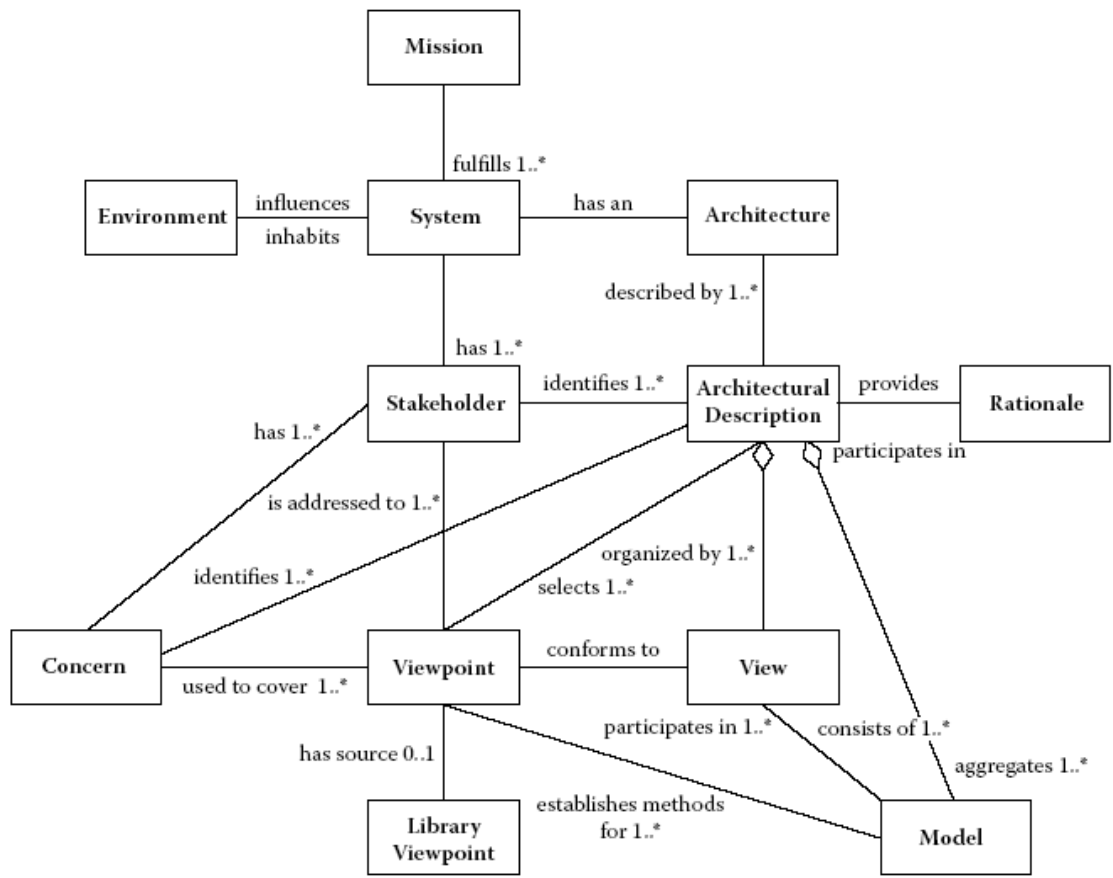

Figure 4: Conceptual framework of IEEE 1471

## 7. THE OPEN GROUP ARCHITECTURE FRAMEWORK (TOGAF)

TOGAF enables corporate architects and stakeholders to design, evaluate, and build flexible enterprise architecture for the organization. The initial versions of TOGAF were based on the Technical Architecture Framework for Information Management (TAFIM), developed by the U.S. Department of Defense (DoD) [11].

There are four types of architectures that are commonly accepted as subsets of overall enterprise architecture, all of which TOGAF is designed to support:

- Business (or business process) architecture: It defines the organization structure, business processes as well as governance.

International Journal of Computer Science & Information Technology (IJCSIT) Vol 3, No 4, August 2011

- Applications architecture: It indicates the base architecture which includes architectural segments along with their interrelationships that conforms to business processes of the organization.
- Data architecture: It describes the data management capabilities grouped to logical as well as physical assets supporting application realization.
- Technology architecture: It is concerned with the infrastructural capabilities which should be considered while implementing and deploying the enterprise solution. As platform independence is a prime issue to be dealt in service composition and availability, it describes the technological alternatives available to male system resources available.

TOGAF has following views and viewpoints for development of enterprise. As mentioned previously, this may be regarded as taxonomy of viewpoints by those organizations that have adopted ANSI/IEEE Std 1471-2000.

- Business Architecture Views, which address the concerns of the users of the system, and describe the flows of business information between people and business processes
- Data Architecture Views, which address the concerns of database designers and database administrators, while identifying and normalizing the database entities of the system.
- Applications Architecture Views, which address the concerns of system and integration engineers responsible for developing and integrating the software components of the system.
- Technology Architecture Views, which address the concerns of acquiring the commercial off-the-shelf (COTS) components that may reduce the cost of software development. The amendments to the components falls into white-box and black-box modifications made to the components. It depends on the suitability of the existing components to identified services to be realized.

## 8. GENERALIZED ENTERPRISE REFERENCE ARCHITECTURE & METHODOLOGY (GERAM)

Previous research, carried out by the AMICE Consortium on CIMOSA, by the GRAI Laboratory on GRAI and GIM, and by the Purdue Consortium on PERA, has produced reference architectures which were meant to be organizing all enterprise integration knowledge and serve as a guide in enterprise integration programs. The IFIP/IFAC Task Force concluded that the architecture derivation should have unique purpose and satisfy the service demands and business needs with a possibility of retainment of service capabilities of previous reference architectures. The recognition of the need to define a generalized architecture is the outcome of the work of the Task Force [12].

The GERA life-cycle for any enterprise consists of different life-cycle phases that define types of activities that are pertinent during the life of the entity. Life-cycle activities encompass activities that span from identification to realization of the enterprise or entity. The activities can be broken into lower level tasks in order to manage the operational effort. Traditional life-cycle management is evident in GERAM methodology with a shift from process components to entities.

International Journal of Computer Science & Information Technology (IJCSIT) Vol 3, No 4, August 2011

- Entity Identification: It describes the entities that constitute the enterprise problem and their limits with possible interactions within the system as well as the external environment. This can be treated as scoping of entities identified..
- Entity Concept: It deals with entity's mission, vision, values, strategies, objectives, operational concepts, policies, business plans which can be used to create entity's knowledge base for further development processes initiation.
- Entity Requirement: The activities needed to develop descriptions of operational requirements of the enterprise entity, its relevant processes and the collection of all their functional, behavioral, informational and capability needs.
- Entity Design: It indicates the process of solution structure and specification of individual components that conforms to the requirements specified.
- Entity Implementation: It describes the effort needed to implement the components identified during the Entity Design step. Reusable components can also be used in concern with cost of modification. If cost of modification of components is higher, components from scratch can be implemented.
- Entity Operation: It deals with deployment of product or service at the customer end. It deals with transition of the solution from source environment to target environment with identification of problems at customer end while using product or services.
- Entity Decommissioning: These activities are needed for future issues like refactoring, reengineering problems associated with the product or services. It emphasizes on the new demands raised to reconsider the problem due to training or design issues.

## 8.1 Modeling Framework of GERA

GERA provides an analysis and modeling framework which is based on the life-cycle approach and indicates following dimensions for defining the scope and content of enterprise modeling.

- *Life-Cycle Dimension:* providing for the controlled modeling process of enterprise entities according to the life-cycle activities.
- *Genericity Dimension:* providing for the controlled particularization (instantiation) process from generic and partial to particular.
- *View Dimension:* providing for the controlled visualization of specific views of the enterprise entity.

### 8.1.1 Entity Model Content Views

Four different model content views define for the user oriented process representation of the enterprise entity descriptions

The *Function View* represents the functions contained in individual business processes and the control applied to each one of them at operational level.

The *Information View* formulates the knowledge base about the entities and the objects identified so as to address the mission and objectives of the enterprise.

The *Resource View* represents hardware, software and human resources required to realize the enterprise solution.

The *Organization View* represents the roles and responsibilities of the people concerned with enterprise development. It also deals with the accountability of human resources in the organization.

### 8.1.2 Entity Purpose Views

- The *Customer Service and Product View* represents the contents relevant to the enterprise entity's operation and to the operation results.

80

- The *Management and Control View* represents the contents relevant to management and control functions necessary to control that part of the enterprise entity that produces products or delivers services for the customer.

### 8.1.3 Entity Implementation View
- The *Human Activities View* represents the set of tasks that are required to be achieved in order to realize the entities identified along with clear description of responsibilities.
- *Automated Activities View* is an indicator of automation effort required to be estimated and delivered to address the technological aspects. This view indicates the tasks that can be automated so as to reduce the manual processing overheads.

### 8.1.4 Entity Physical Manifestation Views
- The *Software View* represents all information resources capable of controlling the execution of the operational tasks in the enterprise
- The *Hardware View* represents the physical resources that are needed to achieve the product functionalities or services at the source and target environments of the enterprise.

## 9. MODEL DRIVEN ARCHITECTURE (MDA)

Model Driven Architecture was introduced by Object Management Group to allow long-term flexibility of implementation, integration, and testing of products and services. Interoperability and platform independence were the two major concerns addressed by MDA. MDA was significantly different approach for specification-based modeling of systems which concentrated on models as a prime issue than objects as in case of object oriented methodologies. MDA introduced model composition and transformation from three levels of models i.e. from Computation-Independent Model (CIM) to Platform-Independent Model (PIM) to Platform-Dependent Model (PSM) based on mapping rules [13]. The core technologies of the OMG MDA are the UML modeling language, the Meta Object Facility (MOF) **Error! Reference source not found.** and the Common Warehouse Metamodel (CWM). Organization of a software system can be represented by structural elements or classes with their interfaces that comprise or form a system and behavior represented by collaboration among these elements. UML is not associated to a process model since it supports the engineering activities ranging from requirements to realization. MOF provides the basis for defining metamodels and model repositories. CWM provides the baseline for data warehousing and data integration. Models are formal specifications of system. A formal specification is consists of syntax, semantics for constructs formulation and usage [14]. The models of the system fall into following categories:
- The conceptual model that captures the system in terms of the domain entities that exist and their association with other system environments.
- The logical view of a system that captures the abstractions indicating the logical separation and boundaries of each identified entity in the conceptual model. It also describes the mechanism through which these entities will interact and form realizable behaviour.
- The physical model of a system describes the software and hardware components that form the system solution space conforming to the specification.

A model can exhibit static structure and defines the universe of discourse. It requires concept mapping from the application domain to a well-formed structure. The analysis classes are transformed to design classes and later to software classes with implementation details of interaction pattern amongst the objects [15]. Dynamic behaviour can be modelled as the life history of one object as it interacts with the rest of the world; the other is the communication patterns of a set of connected objects as they interact to implement behaviour or as the view of an object in isolation is a state machine, a view of an object as it responds to events based on its current state, performs actions as part of its response, and transitions to a new state. Following are the views of "4+1" view architecture:

- Use Case view: It focuses on scenarios indicating the functional requirements which will be used by external entities. This view incorporates analysis level information that dictates static behaviour of the system along with further decomposition of the functionalities.
- Design view: It represents the logical structures which support the requirements expressed in the case view described in terms of classes (and objects) and their behaviour (including interactions between them). It encompasses classes, interfaces, and collaborations that define the vocabulary of a system and supports functional requirements of the system.
- Implementation view: It incorporates physical components that can be grouped into packages indicating realized entities. The basis for these components is analysis and design level classes. The class hierarchy and interaction profile are preserved in this view.
- Process view: It deals with dynamic interaction profile of object including concurrency, time and flow of control. Process view is important in case of real-time applications where synchronization is an important dimension.
- Deployment view: It consists of executables in the form of nodes. Deployment view indicates the resources of system in implementation environment.

The Model-Driven Architecture consists of CIM, PIM, and PSM indicating how they should be used in context of system generation. A viewpoint indicates an aspect or concern of the system which is identified using abstraction principles. A viewpoint model or view of a system is a representation of the domain or partition under consideration. The details of a view can help organize the system elements into realizable components. A platform is a set of functionalities relevant to technology indicating availability of usable services and resources. The platform independence can be achieved by hiding the details of service profiles at software architecture level from the application level by introducing interfaces which can make the resource available from one platform to the other.

- Computation Independent Viewpoint: The computation independent viewpoint focuses on requirements of the system and its structure with environmental needs. It indicates customer, user and stakeholder's perspectives and expectations from system.
- Platform Independent Viewpoint: The platform independent viewpoint focuses on analysis and design models of the system which incorporates the system elements identified and their explicit relationships without adherence to implementation details.
- Platform Specific Viewpoint: The platform specific viewpoint indicates implementation level details of the system elements specific to a particular platform. This can be accomplished by using mapping and transformation rules for migrating from PIM to PSM.

## 10. MEASUREMENT PROCESS

Measurement is the process of describing entities in terms of numbers or symbols. It also indicates the uniqueness property that should be preserved by each identified entity [Fenton 95]. Thus, measurement requires entities (objects of interest), attributes (characteristics of entities) and rules (and scales) for assigning values to the attributes. Measures and metrics are based on measurement scales which can be derived from the rules that we use for assigning values to attributes. Different rules lead to different scales. An ordinal scale permits measured results to be placed in ascending (or descending) order. However, distances between locations on the scale have no meaning. We have used ordinal scale having score values ranging from 0 to 5 as indicated in Figure 5.

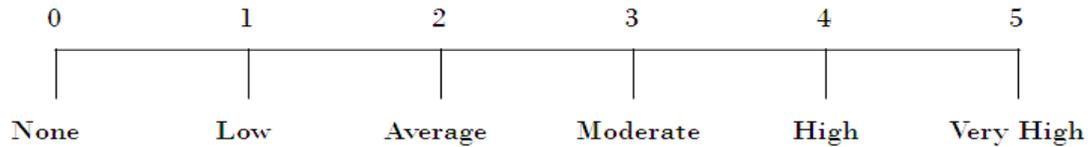

Figure 5: Measurement Scale

## 10.1 COMPARISON BY HIGHER ORDER GOALS

Enterprise integration begins with identification of mission and objectives that directs the business needs of the customer. The enterprise problem then broken into domains that can be implemented and integrated so as to form the enterprise segments [16] [17]. The success of enterprise acceptance depends on customer needs realization and its fulfillment. All enterprises follow a life cycle from their initial concept through a series of stages or phases comprising their development, design, construction, operation and maintenance, refurbishment or obsolescence, and final disposal. Table 2 indicates comparison by higher order goals [18]. Following list indicates higher order goals for Enterprise Architecture:

- Architecture Definition and Understanding – it describes the terminology and guidelines that must be used to define the architecture framework conforming to the needs as stated by the stakeholders identified.
- Architecture Process – it the set of activities performed to attain architecture construction.
- Architecture Evolution Support – it maintains traceability and change profile of system evolution.
- Architecture Analysis – it is a process used to determine the aspects, view and viewpoints that makes up basis of architecture segments.
- Architecture Models – it represents the system in terms of analysis and design models that conforms to standards and specification that guides the development plan.
- Architecture Knowledge Base – it maintains the information base of significant design decisions that directs the enterprise architecture rationale.
- Abstraction – it is an approach to classify the system elements based on similarities and differences. It leads to identification of unique entities of the system.
- Application Architecture – it describes the logical entities and components along with their interaction pattern conforming to identified business needs.
- Architecture Continuum – it is an information base that keeps records of identified architectural segments with appropriate and adequate details so as to realize the architecture. It also encompasses strategies and reference model dictating adoption of architectural styles.
- Architecture Governance – it is the set of processes that guides management and control of the enterprise architectures and other issues related to enterprise-wide level development.

- Architecture Landscape–it deals with identification and management of enterprise assets in accordance with stakeholder needs. It indicates the processes and plans which incorporates strategic and operational profile of the enterprise conforming to stakeholder needs.
- Architecture Verifiability – it provides the set of properties and characteristics that can be checked in order to review the service or product functions.
- Baseline – it is a specification indicating agreed upon properties and characteristics of system that can be examined with current deliverables to estimate its performance. It also serves as important dimension in addressing the changes to be incorporated and its control.
- Business Governance – it indicates the business processes, policies and regulations that need to be practiced while developing the enterprise.
- Capability Architecture – it indicates specification of architectural components with detailed implementation and compositional semantics.
- Data Architecture – it describes the data resources grouped into logical and physical compartments guiding organizational assets.
- Design Tradeoffs – it offers the alternatives for selecting rational design from available choices in order to address the diverse business and technical needs.
- Design Rationale – it indicates the proof of statements for verification and review decisions.
- Data Governance – it indicates the verification mechanisms used to ensure that the data properties and structure has adequate support for transformation and migration.
- Enterprise Continuum – it describes the process of classification of architecture segments and components that makes up the enterprise. It also maintains the catalogue of reference models used; foundation architectures referred leading to custom architectures.
- Environment Management – it indicates the source and target environment in which the system will be operational. It describes the set of resources, facilities and information base that should be made available to deploy enterprise solution.
- Foundation Architecture – it is an architecture of generic services and functions that provides a base for construction of architectural components in question.
- Gap Analysis – it is an indicator of differences between two representations. It is performed to estimate acceptance level of enterprise architecture designed and the baseline considered.
- Metamodel – it is model about model. It specifies the detailed structure and semantics of architectural properties specifications.
- Performance Management– it indicates the post-development activities that needed to be followed to keep track of application performance after deployment.
- Standardization – it indicates whether the determined and accepted standards are met or not.

Table 2: Comparison by higher order goals

| Comparison Parameter | ZF | RM-ODP | FEAF | TOGAF | IEEE 1471 | MDA | GERAM |
|---|---|---|---|---|---|---|---|
| Architecture Definition & | 3 | 5 | 5 | 5 | 5 | 5 | 5 |
| Architecture Process | 0 | 0 | 5 | 5 | 5 | 5 | 5 |
| Architecture Evolution | 0 | 3 | 5 | 5 | 5 | 5 | 5 |
| Architecture Analysis | 5 | 5 | 5 | 5 | 5 | 5 | 5 |
| Architecture Models | 5 | 5 | 5 | 5 | 5 | 5 | 5 |
| Architecture Knowledge Base | 0 | 5 | 5 | 5 | 5 | 5 | 5 |
| Abstraction | 4 | 3 | 4 | 4 | 3 | 4 | 5 |
| Application Architecture | 3 | 2 | 3 | 3 | 3 | 4 | 4 |
| Architecture Continuum | 4 | 4 | 4 | 4 | 3 | 4 | 4 |
| Architecture Governance | 3 | 3 | 4 | 4 | 3 | 4 | 3 |
| Architecture Landscape | 3 | 3 | 3 | 3 | 3 | 4 | 3 |
| Architecture Verifiability | 0 | 3 | 0 | 5 | 5 | 5 | 5 |
| Baseline | 4 | 2 | 3 | 4 | 4 | 3 | 3 |
| Business Governance | 4 | 3 | 3 | 5 | 3 | 4 | 4 |
| Capability Architecture | 3 | 4 | 3 | 4 | 3 | 4 | 3 |
| Data Architecture | 3 | 2 | 4 | 4 | 2 | 3 | 3 |
| Design Tradeoffs | 3 | 3 | 4 | 3 | 4 | 3 | 3 |
| Design Rationale | 3 | 5 | 4 | 5 | 5 | 5 | 5 |
| Data Governance | 3 | 2 | 4 | 4 | 2 | 3 | 3 |
| Enterprise Continuum | 3 | 4 | 4 | 5 | 3 | 4 | 4 |
| Environment Management | 4 | 3 | 3 | 4 | 3 | 4 | 3 |
| Foundation Architecture | 1 | 3 | 2 | 5 | 4 | 4 | 4 |
| Gap Analysis | 3 | 3 | 3 | 5 | 4 | 3 | 4 |
| Metamodel | 3 | 2 | 2 | 5 | 4 | 4 | 4 |
| Performance Management | 2 | 2 | 2 | 4 | 2 | 4 | 4 |
| Standardization | 0 | 5 | 3 | 5 | 5 | 5 | 5 |
| **Total** | **69** | **84** | **92** | **115** | **98** | **108** | **106** |

Table 3: Architecture Definition and Understanding

| Score | Factors indicating Degree of influence |
|---|---|
| 0 | Enterprise Scope and focus is not defined. |
| 1 | The extent of enterprise and architectural effort required to attain the same is defined. |
| 2 | A complete architecture domain description consisting domain information with resource and time constraints is specified. |
| 3 | The level of detail of architecture and architecture effort is determined. |
| 4 | Timing considerations for Architecture Vision realization are indicated. |
| 5 | Target Architecture and Transition Architecture alternatives are defined in order to address the stakeholder objectives in order with increments. |

Table 4: Architecture Process

| Score | Factors indicating Degree of influence |
|---|---|
| 0 | Organizational context for conducting enterprise architecture is not defined. |
| 1 | Organizational context for conducting enterprise architecture is defined and reviewed. |
| 2 | The sponsor stakeholder(s) and other major stakeholders impacted by the business directive are identified to create enterprise architecture and determine their requirements and priorities. |
| 3 | The elements of the enterprise organizations affected by the business directive are identified and scoped with constraints and assumptions. |
| 4 | The framework and detailed methodologies to be used for developing enterprise architectures in the organization concerned are defined. |
| 5 | Target Architecture, infrastructure and supporting tools are selected and implemented. |

Table 5: Architecture Analysis

| Score | Factors indicating Degree of influence |
|---|---|
| 0 | The life cycle management principles and commitments are not defined; hence realistic schedule of architecture development is absent. |
| 1 | Preliminary phases of life cycle are defined and the overall realm of architecture framework is defined and formally stated |
| 2 | The Key Process Areas (KPA) as well as the Key Performance Indicators (KPI) are defined with adherence to the corresponding business processes and drivers . |
| 3 | The Baseline Architecture effort with the relevant stakeholders, and their concerns and objectives is defined. |
| 4 | The development schedule and performance metrics to meet are developed. |
| 5 | Formal approval plan and impact analysis of development cycle is established. |

Table 6: Architecture Verifiability

| Score | Factors indicating Degree of influence |
|---|---|
| 0 | No Architecture Verification iteration exists. |
| 1 | Architecture Context iterations indicating architecture approach, principles, scope, and vision is established. |
| 2 | The iterations required to establish correct and stable architectural information base is established and revised with relevant technical drivers. |
| 3 | Transition Planning iterations supporting formal change adoptions for a defined architecture is established. |
| 4 | Architecture Governance iterations supporting governance of change activity progressing towards a defined Target Architecture is established. |
| 5 | The opportunities and migration planning are traced. |

Table 7: Architecture Governance

| Score | Factors indicating Degree of influence |
|---|---|
| 0 | Governance principles are not established and hence no architecture verification can exist. |
| 1 | All the stakeholders of the enterprise development have agreed upon the processes and deliverables as stated by the stakeholders and recorded by the organization. |
| 2 | All actions implemented and their decision support is available for inspection by authorized organization and provider parties. |
| 3 | All processes, decision-making, and mechanisms used are established so as to minimize or avoid potential conflicts of interest. |
| 4 | Performance metrics and practices to be followed to ensure the architecture enactment policies are determined and monitored. |
| 5 | Stakeholder participation and interaction is determined to monitor progress and performance of architecture development. It principally yields the client and development organization neutral scenario to deploy architectural solution successfully. |

Table 8: Business Governance

| Score | Factors indicating Degree of influence |
|---|---|
| 0 | No description of the Baseline Business Architecture. |
| 1 | Major domain areas and architectural elements are identified formulating the product-level functions and services. Target architecture scope and applicability in corresponding environment are determined. |
| 2 | Reviews of Target Business Architectures and baselines are conducted and examined. |
| 3 | Architecture views and viewpoints are established in accordance with the stakeholder needs and concerns in order to reveal stable architecture segments. |
| 4 | Organization, Goals, Role and Business Service catalogue is developed and standards for each building block from reference model are selected. |
| 5 | Cross check of overall architecture and Architecture Repository mapping is performed. |

Table 9: Standardization

| Score | Factors indicating Degree of influence |
|---|---|
| 0 | Enterprise architecture program is not defined. |
| 1 | The enterprise architecture processes and standards are derived by ad hoc means and are not formal enough to guide the business strategies. |
| 2 | The vision and mission of target enterprise architecture is established with stable and explicit business strategies. |
| 3 | The architecture is well defined and communicated to human resources and management with operation details and responsibilities assigned. It also covers the initial investments to be made along with procurement processes and control. |
| 4 | Enterprise architecture documentation is maintained so as to control and trace the changes incorporated in ongoing development cycle. |
| 5 | Metrics and measures are established and practiced to verify the architecture process. The areas for improvement and optimization of business processes are identified. |

Table 3 to Table 9 indicates the selection criteria on the measurement scale 0 to 5. Architecture Governance, Business Governance and Standardization are the key parameters which determine the applicability of the enterprise architectures depending on the business domain and context identified. Architecture Process and Verification are the other parameters which can be useful in adjudging suitability of the enterprise architecture at the construction and deployment stages. Architecture Analysis depends on baselines and Key Performance Indicators (KPIs).

## 10.2. COMPARISON BY NFR SUPPORT

Requirements are a specification of functions or services that should be accomplished by the system. The requirements are the properties and characteristics possessed by the system along with satisfaction of constraints on them. Requirements vary in intent and in the kinds of properties they represent in terms of *product parameters* and *process parameters*. Product parameters are can be further classified as functional requirements (FR) which indicate what the system should do and affects the performance of the system directly whereas non-functional requirements (NFR) indicate what the system should do and affects the performance of the system indirectly [19] [20].

NFRs are particularly difficult to handle and tend to vary significantly if the goals are expressed ambiguously. Many non-functional requirements have emergent properties. Such requirements

87

cannot be addressed by a single component, but depends for their satisfaction on how all the system components inter-operate. Correctness, consistency, traceability and requirement interaction management are the prime issues to be dealt [21]. Unfortunately, non-functional requirements may be difficult to verify. Non-functional requirements should be quantified. If a non-functional requirement is only expressed qualitatively, it should be further analyzed until it is possible to express it quantitatively. The non-functional requirements mentioned below are quantified on the scale as indicated in the measurement process. Table 10 indicates the comparison by NFR support. Following are the NFRs considered:

- Cohesiveness – It is the degree to which each module in a system does one task and does it well. Cohesion refers to the uniqueness of purpose of the system elements.
- Conceptuality – It represents the concepts in the domain under study. With a conceptual perspective, developers may conceive of what the customer requires, not how. The conceptual level is more abstract than the implementation level, in which the details of how the requirement is to be met are manifested in the code itself.
- Configurability – It describes the ability to organize and control elements of the software configuration. A system's software configuration is defined as the items that comprise all information produced as part of the software process.
- Consistency – It describes two aspects of a system's design and development. Consistency may refer to the use of approaches and techniques describing the system specifications which leads to uniform representations of the system.
- Coupling – It describes the degree to which the modules and components of a given system rely on and interact with other modules and components of that system.
- Diversity – It describes the degree of difference between a system's components and modules. It refers to the degree of difference between data structures and data types throughout a program.
- Extensibility – It involves extending both the design of the system and the software system itself. It describes the degree to which architectural, data, or procedural design can be extended by adding variations to an already stated theme.
- Standardizability – It indicates acceptability and conformance of deliverables against standards. The process standard defines the procedures or operations used in making or achieving a product; the product standard defines what constitutes completeness and acceptability of items that are produced as a result of a process.
- Adaptability – It is defined by the rate at which the software solution can adapt to a new requirement. Adaptability also refers to the degree to which a system may be changed based on a pre-existing system or an unalterable constraint.
- Dependability – It describes the degree to which software performs expected functions and services without failure and acceptable precision.
- Flexibility – It describes the effort required to modify an operational program or system. A software system may be required to be flexible if there will be known a change in its operating environment after it has been deployed and is in normal operation.
- Maintainability – It describes the effort required to locate and fix an error in a program. It the ease with which a program can be corrected if an error is encountered, adapted if its environment changes, or enhanced if the customer desires a change in requirements.
- Maturity – It describes the degree to which a software system is mature. A system is said to be mature when it has attained a final, desired state of full development.
- Portability – It describes the ease with which the software can be transposed from one environment to another.

- Scalability – It refers to the ease with which a system may be made smaller or larger, although most of the time, increasing the system's size is the concern, not reducing it.
- Robustness – It describes the degree to which a program or system can recover gracefully whenever a failure occurs. It also describes the time it takes the system to restart after experiencing system failure.
- Security – It describes the mechanisms that detect the possible threats to programs and data. It may also refer to the probability that the attack of a specific type will be repelled.
- Compatibility – It describes the ability of two or more systems to exchange information. When a system is being deployed to replace an earlier version of that system, it is imperative that it be compatible with everything that it is replacing is compatible with.
- Inter-operability – It is defined as the ability of the systems to exchange the services with agreed protocols and architectural support at both the ends.
- Usability – It describes the effort required to learn and handle the services or product functions over a period of time.

Table 10: Comparison by NFR support

| Comparison Parameter | ZF | RM-ODP | FEAF | TOGAF | IEEE 1471 | MDA | GERAM |
|---|---|---|---|---|---|---|---|
| Adaptability | 4 | 4 | 3 | 5 | 4 | 5 | 4 |
| Compatibility | 3 | 4 | 3 | 5 | 3 | 4 | 4 |
| Cohesiveness | 3 | 3 | 4 | 4 | 4 | 4 | 4 |
| Conceptuality | 4 | 4 | 4 | 5 | 4 | 4 | 4 |
| Configurability | 2 | 4 | 4 | 4 | 4 | 4 | 4 |
| Consistency | 3 | 3 | 4 | 5 | 4 | 4 | 4 |
| Coupling | 3 | 3 | 4 | 5 | 4 | 4 | 4 |
| Diversity | 3 | 3 | 3 | 5 | 3 | 5 | 3 |
| Dependability | 3 | 4 | 4 | 4 | 4 | 4 | 4 |
| Extensibility | 3 | 3 | 4 | 4 | 3 | 4 | 4 |
| Flexibility | 3 | 4 | 3 | 5 | 4 | 4 | 4 |
| Inter-operability | 3 | 3 | 3 | 5 | 3 | 5 | 3 |
| Maintainability | 3 | 4 | 4 | 4 | 3 | 4 | 3 |
| Maturity | 3 | 3 | 3 | 4 | 4 | 4 | 4 |
| Portability | 2 | 4 | 3 | 4 | 3 | 4 | 3 |
| Robustness | 3 | 4 | 4 | 4 | 3 | 4 | 4 |
| Scalability | 3 | 3 | 4 | 4 | 4 | 4 | 4 |
| Security | 2 | 3 | 4 | 4 | 3 | 4 | 3 |
| Standardizability | 3 | 3 | 4 | 5 | 4 | 4 | 3 |
| Usability | 4 | 3 | 3 | 5 | 3 | 4 | 3 |
| Total | 60 | 69 | 72 | 90 | 71 | 83 | 73 |

## 10.3. COMPARISON BY INPUTS AND OUTCOMES

Business drivers, Technology inputs, and Business requirements focus on the problem issues in view of the stakeholders. The context and relevance of the problem scenario can be further broken into various model supports as indicated in Table 11. The process enablers as well as process measures are key areas determining sustainability and stability of the enterprise solution.

Table 11: Comparison by Inputs and Outcomes

| Comparison Parameter | ZF | RM-ODP | FEAF | TOGAF | IEEE 1471 | MDA | GERAM |
|---|---|---|---|---|---|---|---|
| Business Drivers | 3 | 3 | 5 | 5 | 3 | 5 | 3 |
| Technology Inputs | 0 | 3 | 5 | 5 | 5 | 5 | 4 |
| Business Requirements | 5 | 5 | 5 | 5 | 3 | 5 | 3 |
| Information System | 3 | 5 | 5 | 5 | 5 | 4 | 4 |
| Existing Architecture | 3 | 5 | 5 | 5 | 5 | 4 | 5 |
| Business Model Support | 5 | 5 | 5 | 5 | 3 | 5 | 3 |
| System Model Support | 5 | 5 | 5 | 5 | 5 | 5 | 4 |
| Information Model Support | 5 | 5 | 5 | 5 | 5 | 4 | 4 |
| Computation Model Support | 5 | 5 | 5 | 5 | 5 | 5 | 4 |
| Software Configuration | 0 | 3 | 0 | 5 | 4 | 4 | 4 |
| Software Process Incorporation | 4 | 4 | 4 | 5 | 3 | 4 | 3 |
| Implementation Model | 3 | 4 | 4 | 4 | 3 | 4 | 4 |
| Platform | 4 | 5 | 4 | 4 | 3 | 5 | 4 |
| **Total** | 45 | 57 | 57 | 63 | 52 | 59 | 49 |

Figure 6 indicates the consolidated chart representing the enterprise architecture suitability depending on higher order goals, NFR support and input-outcomes.

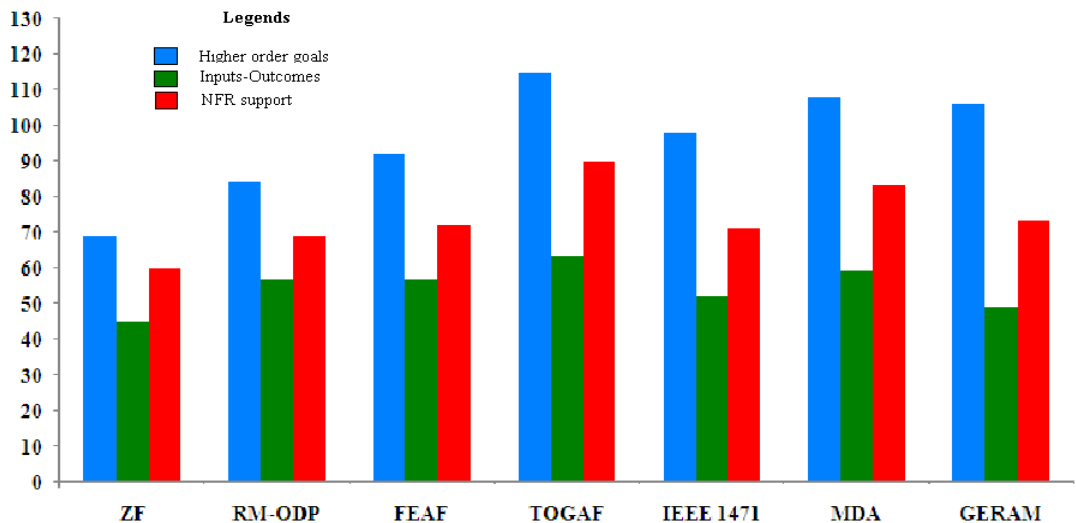

Figure 6: Consolidated Comparison Chart

## 11. CONCLUSIONS

The paper covers a broad discussion of major enterprise architecture methodologies. The enterprises can be categorized into small-sized, medium-sized and large-sized enterprises depending on the range of problem issues, business requirements, and organization portfolio. It is significantly difficult to decide on selecting a specific enterprise architecture methodology due to the changes that drives the enhancement scenario for these methodologies. Every system development effort is constrained by the time, scope and cost triplet. The relationship between scope and performance has to be established at the time of system conceptualization so that realistic solution with required fitness criteria can be developed.

The paper proposed an ordinal scale based measurement process for measuring enterprise architecture methodologies in terms of higher order goals, NFR support and input-outcomes. It can be observed that TOGAF and MDA are the most successful methodologies in addressing the issues indicated due to incorporation of views and viewpoints. Business, Architecture, Technology and Data governance are also the key areas which indicate the rationale and applicability of the methodologies. However, the fundamental methodology proposed by Zachman Framework is nearly adopted and considered by every descendant methodology development effort.

The paper focused on the criticality of addressing NFR issues. NFR properties are the abilities that the system should possess that ensure required quality and performance has been met at product or service level. We have considered major NFRs that can impact the selection of enterprise architecture methodologies. It can be observed that TOGAF, MDA, GERAM and IEEE 1472-2000 are in a comparable range in this context. The paper also suggests that there cannot be a radical shift from one methodology to the other since methodology mapping must be discovered before doing so. Finally, the selection of any enterprise architecture methodology will depend on organization culture, mission, principal investment at the initial phase and adherence to the architecture principles.

**Authors**

Mahesh. R. Dube (mrdube@rediffmail.com) has completed ME in Computer Science and Engineering from Vishwakarma Institute of Technology, Pune, India in 2006. Currently, he is working as Assistant Professor at Department of Computer Engineering, Vishwakarma Institute of Technology, Pune. He is perusing his Ph.D. from Solapur University, Maharashtra, India, in integrated modeling and design. His areas of interest are Software Engineering, Modeling and Design, Software Testing and Design Patterns.

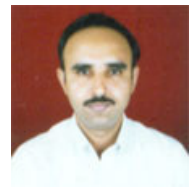

Shantanu K. Dixit (dixitsk1@yahoo.com) has received Ph.D. in Electronics from Shivaji University, Kolhapur in 2002. Currently, he is working as Professor and Head, Department of Electronics and Telecommunication at Walchand Institute of Technology, Solapur, Maharashtra, India. His research areas are Robotics, Modeling and Simulation, Communication Engineering and Microcontroller based design.

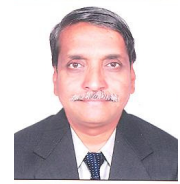